\documentclass[applsci,article,accept,moreauthors,pdftex]{mdpi}

\usepackage{soul}
\usepackage{booktabs}
\usepackage{multirow}
\usepackage{microtype}
\usepackage{subfigure}
\makeatletter
\renewcommand{\@thesubfigure}{\normalsize(\textbf{\alph{subfigure}})}
\makeatother
\usepackage{hhline}

\usepackage{amsmath,amssymb,amsfonts}
\usepackage[ruled,vlined]{algorithm2e}
\SetKwRepeat{Do}{do}{while}%
\usepackage{graphicx}
\usepackage{textcomp}
\usepackage{amsmath}
\usepackage{mathtools}
\usepackage[table]{xcolor}
\DeclareMathOperator*{\argmin}{arg\,min}

\firstpage{1}
\pubvolume{xx}
\issuenum{1}
\articlenumber{5}
\pubyear{2020}
\copyrightyear{2020}
\history{Received: 11 August 2020; Accepted: 11 September 2020; Published: date}
\updates{yes} % If there is an update available, un-comment this line

\Title{Neural Network Based Country Wise Risk Prediction of COVID-19}
\Author{Ratnabali Pal $^{1}$\orcidA{}, Arif Ahmed Sekh $^{2}$\orcidA{}, Samarjit Kar $^{3}$ and Dilip K. Prasad $^{1,}$*\orcidA{}}
\AuthorNames{Firstname Lastname, Firstname Lastname and Firstname Lastname}

\address{%
$^{1}$ \quad Department of Computer Science, UiT The Arctic University of Norway, Troms{\o} 9019, Norway;  pal.ratnabali@gmail.com (R.P.)\\ %MDPI: Please add post code, city and country. (or zip code in the US). %MDPI: Please add the department/school/faculty/campus.  The format of affiliations: Department, University/Company, post/zip code City, Country.
$^{2}$ \quad Department of Physics and Technology, UiT The Arctic University of Norway, Troms{\o} 9019, Norway; skarifahmed@gmail.com (A.A.S.)\\ %MDPI: Please add post code, city and country. (or zip code in the US). %MDPI: Please add the department/school/faculty/campus.   The format of affiliations: Department, University/Company, post/zip code City, Country.
$^{3}$ \quad Department of Mathematics, National Institute of Technology Durgapur, Durgapur 713209, India; samarjit.kar@maths.nitdgp.ac.in (S.K.)} %MDPI: Please add post code, city and country. (or zip code in the US). %MDPI: Please add the department/school/faculty/campus.   The format of affiliations: Department, University/Company, post/zip code City, Country.

\corres{Correspondence: dilip.prasad@uit.no} %MDPI:It's different from susy, please confirm if it is right. Ans: It is right

\abstract{The recent worldwide outbreak of the novel coronavirus (COVID-19) has opened up new challenges to the research community. Artificial intelligence (AI) driven methods can be useful to predict the parameters, risks, and effects of such an epidemic. Such predictions can be helpful to control and prevent the spread of such diseases. The main challenges of applying AI is the small volume of data and the uncertain nature. Here, we propose a shallow long short-term memory (LSTM) based neural network to predict the risk category of a country. We have used a Bayesian optimization framework to optimize and automatically design country-specific networks. The results show that the proposed pipeline outperforms state-of-the-art methods for data of 180 countries and can be a useful tool for such risk categorization. We have also experimented with the trend data and weather data combined for the prediction. The outcome shows that the weather does not have a significant role. The tool can be used to predict long-duration outbreak of such an epidemic such that we can take preventive steps earlier.}

\keyword{COVID-19; trend prediction; optimized neural network}

\begin{document}
%%%%%%%%%%%%%%%%%%%%%%%%%%%%%%%%%%%%%%%%%%

%%%%%%%%%%%%%%%%%%%%%%%%%%%%%%%%%%%%%%%%%%
%\setcounter{section}{-1} %% Remove this when starting to work on the template.
\section{Introduction}
\label{sec:introduction}
\begin{figure}[H]
    \centering
    \includegraphics[scale=0.68]{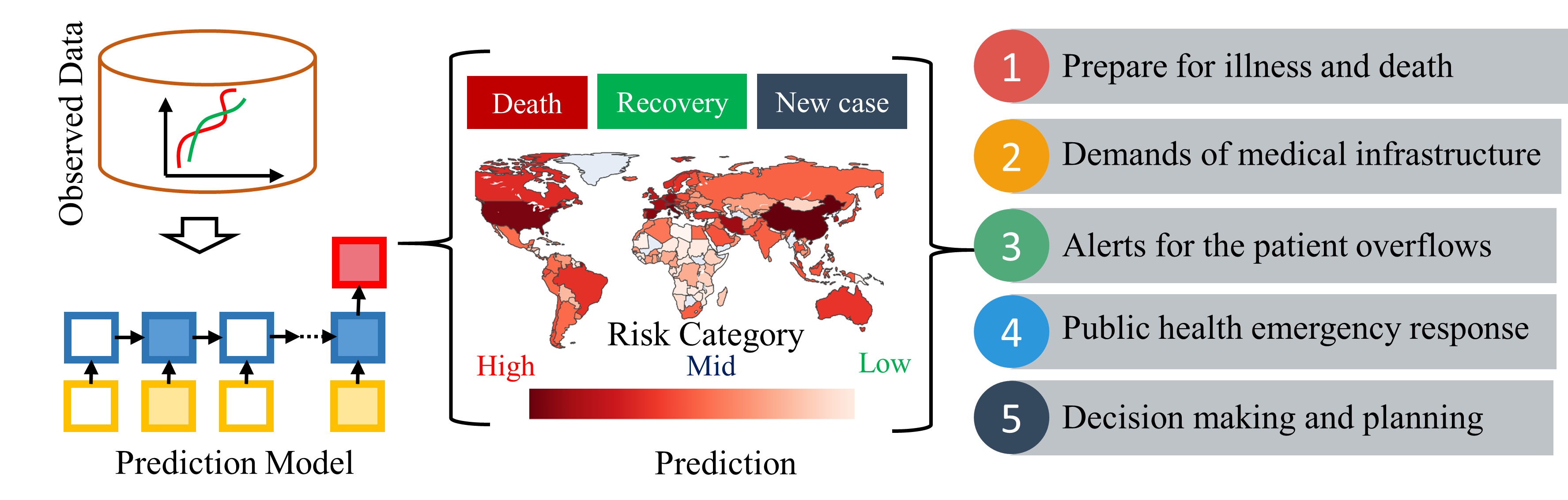}
    \caption{{Artificial intelligence (AI) method for trend prediction and applications of such prediction.}}
    \label{fig:teaser}
\end{figure}

The novel coronavirus (COVID-19) hit our blue planet and became an ongoing global pandemic~\cite{wu2020new}. In a little over six months since the virus was first spotted in mainland China, it~has spread to more than 180 countries, infected more than 18.4 million people, and taken more than 692,000 lives as reported in the first week of August 2020.
As governments and health organizations scramble to contain the spread of coronavirus, they need all the help they can get, including from artificial intelligence (AI). Though the current AI technologies are far from replicating human intelligence, they~are proving to be helpful in tracking the outbreak, diagnosing patients, disinfecting areas, and~speeding up the process of finding a cure for COVID-19.
{Forecasting is a collection of quantitative, probabilistic statements based on historical observation. It is a process of predicting unobserved events and trends. Population~health monitoring and forecasting, including epidemiological outbreaks may not have any clinical utility but can be useful tool for planning, decision making and prevention (see Figure \ref{fig:teaser}).

The early prediction of epidemics will benefit governments and health-care departments to enable a timely response to outbreaks. It will minimize the impact and ensure the use of resources in a planed manner. For many contagious diseases, location specific prediction of the trend is useful to minimize the risk of spread at community level by limiting social gatherings and imposing travel restrictions. Similarly, strategy makers utilize infectious disease forecasts towards prepare medical and economic preparedness. The forecasting of social impact~\cite{cheongintroducing} is also useful for long-term strategy makers.}

{In the last few years, AI methods have been successfully applied to various predictive tasks such as stock value~\cite{akita2016deep}, sales~\cite{ali2018crm} and weather~\cite{xiao2018data} prediction, as well as predicting epidemic spread~\cite{lu2019epidemic}. Prediction techniques also proved to be useful in many healthcare applications. Lin et al.~\cite{lin2020leveraging} shows how the prediction of ambulance demand provides great value to emergency service providers.}

In this paper, we have proposed an AI-guided method to predict long-term country-specific risk of coronavirus. The primary challenges of this problem are:

Small dataset: Majority of machine learning (ML) algorithms demand a large volume of data for training. Notably, the COVID-19 dataset is less than a year-long and it is difficult to design accurate AI methods to train on such small volume of data.
%Is bold neccessary? the same to following format like this. Ans: Removed

Uncertain data: The virus is quite new to the researchers and the majority of the parameters that can be used to predict the outbreak and risk factors are unknown. It is observed that the trend is also different in different countries. Hence, a generic AI tool may not be suitable for tracking all trends. It is also noted state-of-the-art deep neural networks fail because of the uncertainty in the data. This observation encourages us to design shallow and country-specific optimized neural networks.

Data fusion: In many articles, it is claimed that the weather has a role in the outbreak of the virus. Most of the research works are shown in a country-specific manner~\cite{tosepu2020correlation,gupta2020effect,csahin2020impact}. A systematic analysis of worldwide different weather parameters and the outbreak status can be useful for better understanding the relation.

Here, we have proposed to use the local data trend with a shallow Long Short-Term Memory (LSTM) based neural network combined with a fuzzy rule based system to predict long term risk of a country (Figure~\ref{fig:teaser}). The country-specific neural networks are optimized using Bayesian optimization.

%Next, we discuss the related works and gaps bridge by the proposed method.

\section{Related Works}
We note three communities of the related work: (A) AI in epidemic research, (B) research works on COVID-19, and (C) multivariate regression in AI. These are discussed below:

(A) AI based epidemic researches: Real-time epidemic-forecasting attracts several researchers due to the emerging applicability of the method. Jia et al.~\cite{jia2019predicting} proposed a neural network for predicting the outbreak of hand-foot-mouth diseases. Hamer et al.~\cite{hamer2020spatio} used ML algorithms to predict spatio-temporal epidemic spread of pathological diseases. AI tools for predicting outbreak in cardiovascular diseases~\cite{mezzatesta2019machine,jhuo2019trend}, influenza~\cite{kumar2020outbreak}, and epidemic diarrhea~\cite{machado2019identifying} is also proposed. A review of the application of AI for such a prediction is reported in~\cite{philemon2019review}. A collective learning based approach~\cite{abdulkareem2020risk} is proposed to identify individual risk. In the last few years, machine learning analysis was used to predict epidemiological characteristics of the Ebola virus (EBOV) outbreak in West Africa~\cite{forna2019case} and the risk of Nipah virus~\cite{dallatomasina2015ebola}. Plowright et al.~\cite{plowright2019prioritizing} proposed a surveillance method to monitor Nipah virus in India. Recently, Seetah et al.~\cite{seetah2020archaeology} proposed a method for predicting future Rift Valley fever virus outbreaks. The majority of the algorithms use a combined decision-making application using statistical and machine learning methods to predict future growth based on past incident data.

(B) Researches on COVID-19: The recent COVID-19 outbreak has motivated many researchers to help and find a way to recover from the pandemic. Rao et al.~\cite{rao2020identification} proposed methods to detect COVID-19 patients using a mobile phone. Yan et al.~\cite{yan2020prediction} built a predictive model to identify early detection of high-risk patients before their health status is transformed from mild to critically ill. In~recent times, numerous research articles have been published on epidemic prediction of the coronavirus pandemic~\cite{peng2020epidemic,zhao2020preliminary,chen2020data,li2020scaling,hilton2020estimation,kastner2020viewing,jia2020prediction,zhao2020tracking,zeng2020predictions,buizza2020probabilistic}. Researchers designed new paradigms of AI-driven tools~\cite{fong2020finding,santosh2020ai} that combine ML algorithms and different modalities of data. An improved adaptive neuro-fuzzy inference system (ANFIS) methodology is proposed in~\cite{al2020optimization}. The algorithm is based on an enhanced flower pollination algorithm (FPA) by using the salp swarm algorithm (SSA) to estimate confirmed cases in the next 10 days. Li et al.~\cite{li2020covid} developed a regression model to calculate the exponential growth of COVID-19 infection based on the total number of daily diagnoses cases outside China. Analysts in ~\cite{araujo2020spread} obtained projections from 10 familiar machine learning and statistical	
ecological niche models against the large-scale climatology variation.

(C) Multivariate Regression in AI: The key point in time series study~\cite{dong2019partial} is forecasting. Time~series analysis for business prediction helps to forecast the probable future values of a practical field in the industry~\cite{moews2019lagged,thomas2019time,lorenzo2019some,bandara2019sales}. The method is also applicable in the health domain to predict the health condition of a person on the last diagnosis data\cite{cui2019prediction}. The method uses a feature attention mechanism to predict future health risks. {Oh et al. \cite{oh2018automated} use a combination of convolutional neural network (CNN) and LSTM for automated diagnosis of arrhythmia. The input electrocardiogram signal is processed using CNN and processed using LSTM to handle variable length signal. A multiple regression predictive model~\cite{ho2019forecasting} was used to predict patient volume in the hospital emergency departments. The authors used Google trend for forecasting.}
Other health areas such as antibiotic resistance outbreaks~\cite{jimenez2020feature} and influenza outbreaks~\cite{tapak2019comparative,su2019forecasting} utilized multivariate regression models. Different algorithms such as deep neural network~\cite{ochodek2020deep,hu2020efficient}, long short-term memory model (LSTM)~\cite{wen2019real} and gated recurrent unit (GRU)-based model~\cite{yuan2019novel} have been successfully applied in various forecasts. The methods rely on specific-less estimation error and running time on data sets with characteristics of multivariate, sequential and time-series data.

Gap bridged by our method: The main challenge of predicting the long term risk of a country is solved by designing dynamic shallow recurrent neural network (RNN) which is optimized for an individual country, and combining fuzzy rules for inference. It is reported in~\cite{santosh2020ai} that designing a custom network based on input data is a suitable solution. This observation inspired us to design an optimized network for each country. The problem of insufficient data is solved by choosing an optimized shallow network and the problem of predicting local trends is solved by optimizing the neural networks for individual countries. This introduces a new way to predict an epidemic outbreak and correlate with the risk of a country. In many research works~\cite{tosepu2020correlation,gupta2020effect,csahin2020impact}, correlation of weather with the virus spread is indicated. We have analyzed using Spearman's rank correlation coefficient analysis and ordinary least squares (OLS) regression and found that the prediction of new cases, recovery,~and~death does not depend on the weather. The proposed neural models perform similar (or better in some cases) without the weather data.

\section{Proposed Model}
The proposed framework consists of four modules as shown in Figure ~\ref{fig:proposed}. The modules are (1) search space definition module, (2) network search module, (3) local trend prediction, and (4) a fuzzy rule-based risk assessment module. We first discuss the background of RNN and then these modules~below: %MDPI:Is the bold necessary? Ans: removed
\begin{figure}[H]
    \centering
    \includegraphics[width=\linewidth]{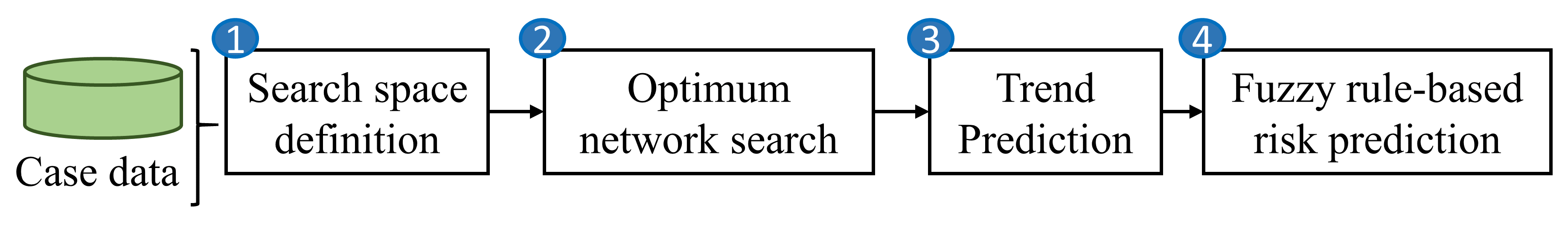}
    \caption{Modules of the proposed framework.}
    \label{fig:proposed}
\end{figure}
%------------------------------------------------------------%
\subsection{Background}
We propose to use a shallow long short-term memory (LSTM) with a few layers. LSTM is also a variation of RNN like GRU. Fundamentally, an RNN handles the sequence by having a recurrent hidden state whose activation at each time is dependent on the state at the previous time. \mbox{Formally, for a} set of input $x = (x_1, x_2, · · · , x_T )$, the RNN estimates its hidden state $h_t$ by
\begin{equation}
    h_t = \begin{cases}
o &\text{$t=0$}\\
\nu (h_{t-1},x_t) &\text{otherwise}
\end{cases}
\end{equation}
where $\nu$ is the nonlinear function. %The LSTM have an output $y = (y_1, y_2, . . . , y_T )$.
The hidden states are updated by
\begin{equation}
    h_t = g (W x_t + U h_{t-1})
\end{equation}
where $g$ is a bounded function. A general RNN estimates the conditional probability of each input state as
\begin{equation}
    p(x_t  x_1, . . . , x_{t-1}) =g(h_t)
\end{equation}

LSTM is adaptive and estimates dependencies of different time scales. The commonly used RNN variations such as LSTM use gate and memory cells for sequence prediction. Initially, LSTM initiates with a forget gate layer $(f_t)$ that uses a sigmoid function combined with the previous hidden layer $(h_{t-1})$ and the current input $(x_t)$ as:
\begin{equation}
    f_t=\sigma (W_f \cdot [h_{t-1},x_t]+b_f)
\end{equation}
where $W$ is weight and $b_f$ is the bias. A hypertangent layer ($\widetilde{C}$) is represented by a tanh cell as:
\begin{equation}
    \widetilde{C}=tanh(W_C \cdot [h_{t-1},x_t]+b_C)
\end{equation}

\begin{figure}[H]
    \centering
    \includegraphics{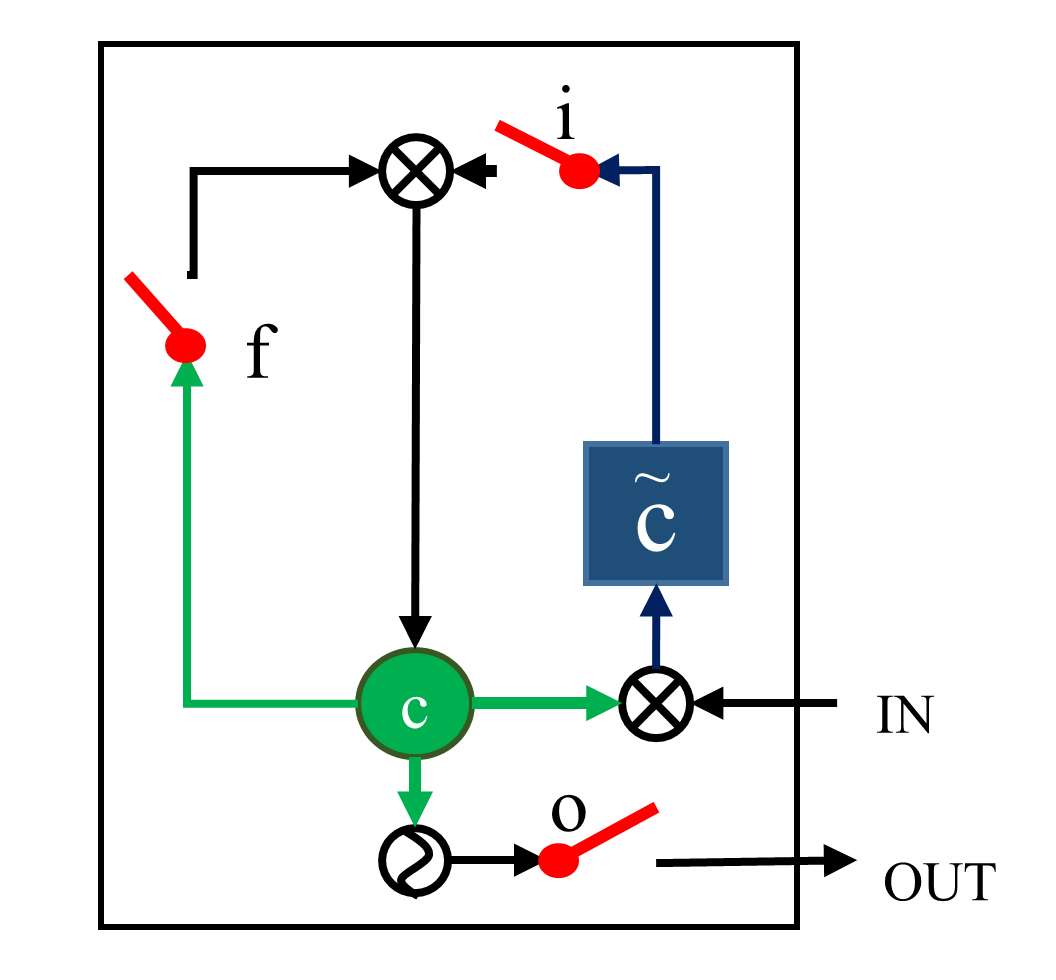}
    \caption{The architecture of LSTM module. Here,
$i$ is the input gate, $f$ is the forget gate and $o$ is the output gate. $c$ is the cell state and $\widetilde{C}$ is the update cell.}
    \label{fig:lstm}
\end{figure}

This information is passed to the next cell $C_t$ as:
\begin{equation}
    C_t=f_t * C_{t-1}+i_t * \widetilde{C}_t
\end{equation}
where $i_t$ also a sigmoid function. Finally, this information passed to the next hidden layers as:
\begin{equation}
    h_t=o_t*tanh(C_t)
\end{equation}
where $o_t$ is also a sigmoid function known as the output gate. The graphical representation of LSTM is presented in Figure~\ref{fig:lstm}.
We have used a similar structure of LSTM module as the building blocks of the proposed system.
%------------------------------------------------------------%
\subsection{Search Space Definition} %We hypothesize that a neural network with a fixed parameter setup is not suitable for each country.
Each country has a different trend and demands a neural network with different parameter setup. The input data is fixed for all countries that contain three main concerns for the risk categorization. Number of cases ($\kappa$), number of deaths ($\delta$), and number of recoveries ($\rho$). The number of active cases ($\alpha$) is calculated by $\kappa-(\rho + \delta)$. The learning algorithm (here LSTM) is defined by a set of parameters. Let $\{\phi_1,\phi_2,...,\phi_n\}$ be a set of hyperparameters of the learning algorithm and $\mu_1,\mu_2,...,\mu_m$ be domains of the parameters, known as the search space. Table~\ref{tab:param} summarized the parameters and the search space used in our method.
\begin{table}[H]
\centering
\caption{Parameters used in optimum network search for novel coronavirus (COVID-19).}
\begin{tabular}{cccc}
\toprule
\textbf{Parameter} & \textbf{Description}                & \textbf{Distribution/Selection} & \textbf{Values}       \\ \midrule
Learning rate      & Minimum learning rate               & Log uniform                     & 1 $\times$ $10^{-1}$  to 1 $\times$ 10$^{-7}$          \\ \midrule
Hidden layers      & Number of layers in the network     & Discrete numeric                & 1 to 20               \\ \midrule
Hidden state       & Number of memory cell in each layer & Discrete numeric                & 1 to 200              \\ \midrule
Activation         & Activation in each layer            & Category                        & \{ReLU,sigmoid,tanh\} \\ \midrule
Batch size         & Batch size during training          & Discrete numeric                & 2 to 10               \\ \midrule
Dropout           & Dropout size before dense layer     & Log uniform                     & 0 to 0.5              \\ \bottomrule
\end{tabular}

\label{tab:param}
\end{table}
%------------------------------------------------------------%
\subsection{Network Search} Let $\phi_1,\phi_2,...\phi_n$ be different hyperparameters of a learning algorithm and $\mu_1,\mu_2,...,\mu_n$ be domains of the parameters as defined earlier. The dataset ($D$) is divided into train ($D_{train}$) and test ($D_{test}$). The hyperparameter space is $\theta = \mu_1 \times \mu_2... \times \mu_n$. Training data is trained on $\phi \in \theta$. The test error $E(\phi,D_{train},D_{test})$ is the error on $D_{test}$ of the parameter $\phi$. The hyperparameter is optimized for a given dataset
($D$) by minimizing:
\begin{equation}
    f^D(\phi)=E(\phi,D_{train},D_{test})
\end{equation}

We have considered root-mean-square error (RMSE) on validation set to chose best architecture. Hence, the problem can be defined as:
\begin{equation}
    \phi ^ * = \argmin_{\phi \in \theta}(f^D(\phi))
\end{equation}

In general, the problem of hyper-parameter search can be very expensive as we need to train and evaluate the dataset for each combination of parameters. Searching algorithms such as random search and grid search are better than manual setup but computationally expensive when we have a large volume dataset and a wide hyper-parameter search space. These methods do not consider the previous outcome to choose the next set of parameters, hence they spend most of the time evaluating bad parameters. In our case, the RMSE of a set of parameters ($ f^D(\phi_{next}$) is estimated by the conditional probability $P(f^D(\phi_{next})f^D(\phi_{previous}))$. The method selects the set of hyper-parameters that performs the best according to the probability.
First, individual COVID-19 trend is used to automatically design the desired neural network. Next, the network optimized for an individual country is used to predict the number of cases ($\kappa$), the number of deaths ($\delta$), and the number of recoveries ($\rho$). These data are used in the next module to decide the risk of the country.

%------------------------------------------------------------------------%
\subsection{Fuzzy Rule-Based Risk Categorization}
The prediction of $\delta$, $\kappa$, and $\rho$ is used to predict the risk for the country. We define 3 categories of risks (1) high risk (HR), (2) medium risk (MR), (3) recovering (RE). First, we calculate the death rate, the rate of new cases, and the recovery rate as: %MDPI: Is the bold necessary? If not, please remove the format. Ans; Removed
\begin{equation}
    \textnormal{\textit{death rate}}= \frac{\kappa}{\delta}
\end{equation}
\begin{equation}
    \textnormal{\textit{case rate}}=\frac{\textnormal{\textit{total population}}}{\alpha}
\end{equation}
\begin{equation}
    \textnormal{\textit{recovery rate}}=\frac{\alpha}{\rho}
\end{equation}

\begin{figure}[H]
    \centering
    \includegraphics[width=\linewidth]{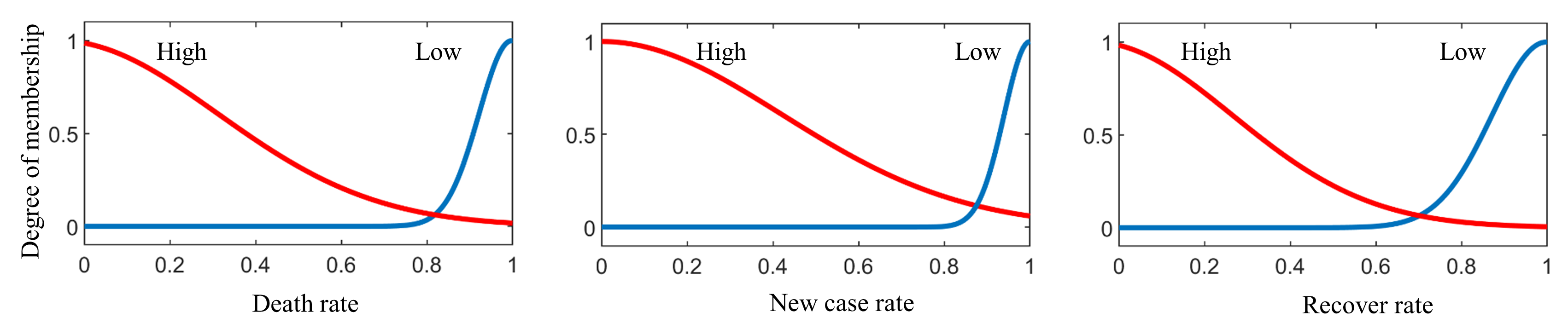}
    \caption{Fuzzy membership functions for death rate, case rate, and recovery rate.}
    \label{fig:fuzzy}
\end{figure}

Next, three Gaussian fuzzy membership functions are defined to represent the risk measurement of these parameters as shown in Figure~\ref{fig:fuzzy}. The final class of the risk is estimated my imposing rules defined in Table~\ref{tab:fuzzy}.
\unskip
\begin{table}[H]
\centering
\caption{Fuzzy rules to estimate the risk factor of a country}
\label{tab:fuzzy}
\begin{tabular}{cccc}
\toprule
\textbf{Death rate} & \textbf{Case rate} & \textbf{Recovery rate} & \textbf{Decision}                  \\ \midrule
High                & High               & Low                   & HR \\ \midrule
Low                 & High               & Low                   & HR \\ \midrule
High                & High               & High                  & HR \\ \midrule
Low                 & High               & High                  & HR \\ \midrule
High                & Low                & High                  & MR \\ \midrule
High                & Low                & Low                   & MR \\ \midrule
Low                 & Low                & Low                   & MR \\ \midrule
Low                 & Low                & High                  & RE \\ \bottomrule
\end{tabular}
\end{table}
 %%MDPI: is color neccessary? if it is, please provide explanations under the table, the same to Tables 3 and 4.
 %%Ans: Removed
%------------------------------------------------------------------------------------------------%

\section{Results and Discussion}
We conducted various experiments using different baseline algorithms and our proposed method. We have extensively analyzed the results from different perspectives. First, we present the effectiveness of the feature selection method. Next, we discuss the results of proposed network optimization, and~we compare the method with the baselines. Finally, we conclude the article with our findings.
\subsection{Dataset} We used the dataset (https://github.com/datasets/covid-19) that included date, \mbox{country, the number} of confirmed cases, the number of recovered cases, and the total number of deaths. We combined this data with weather data~(https://darksky.net/) consisting of humidity, dew, ozone, perception, maximum temperature, minimum temperature, and UV for analyzing the effect of weather. We considered mean and standard deviation over different cities of a country. The data spanned the duration 22-01-2020 to 02-08-2020. %Footnote is not permitted in this journal, so we have moved it into the text, please confirm. Ans: Fine
\subsection{Network Optimization}
\begin{figure}[H]
    \centering
    \includegraphics[width=\linewidth]{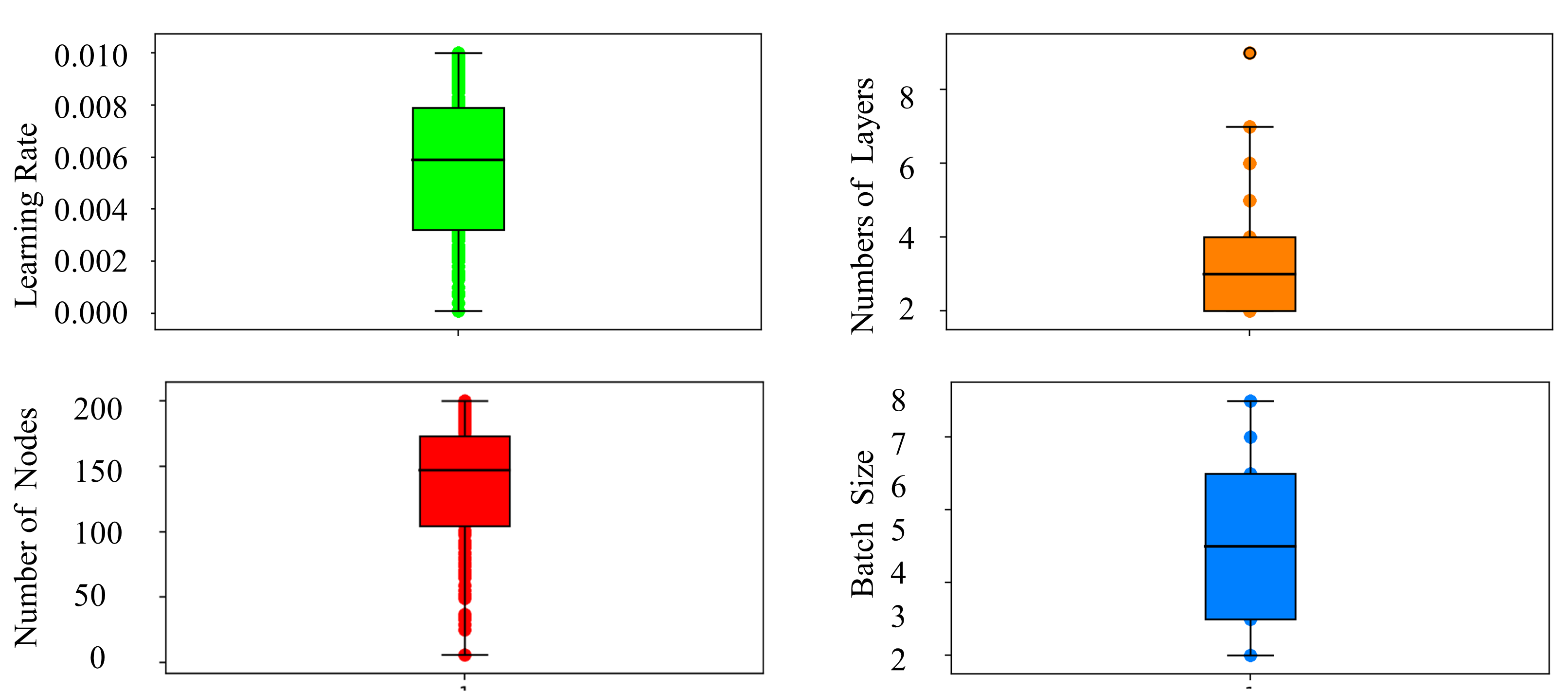}
    \caption{Distribution of the parameters of the optimized for 180 country specific Long Short-Term Memories (LSTMs).}
  \label{fig:opt_param}
  \vspace{3mm}
\end{figure}

We used 300 iterations with Gaussian noise with variance 0.01 added to the data. The~data of the last 25 days were used for validation and the rest were used for training. Each network generated by Bayesian optimization was trained using a maximum of 5000 iterations. We used 100 epoch delay on validation loss for early stopping. We used 300 iterations during optimization. During~optimization, RMSE was minimized over the validation set. The data of each country were individually used to generate the country-specific optimized network. It was observed that majority of the optimal networks comprised of only a few layers and hidden units with ReLU activation. The~distribution of the parameters over all the generated networks is shown in Figure~\ref{fig:opt_param}. The dropout was chosen as zero most of the time.

\begin{figure}[H]
    \centering
    \includegraphics[width=\linewidth]{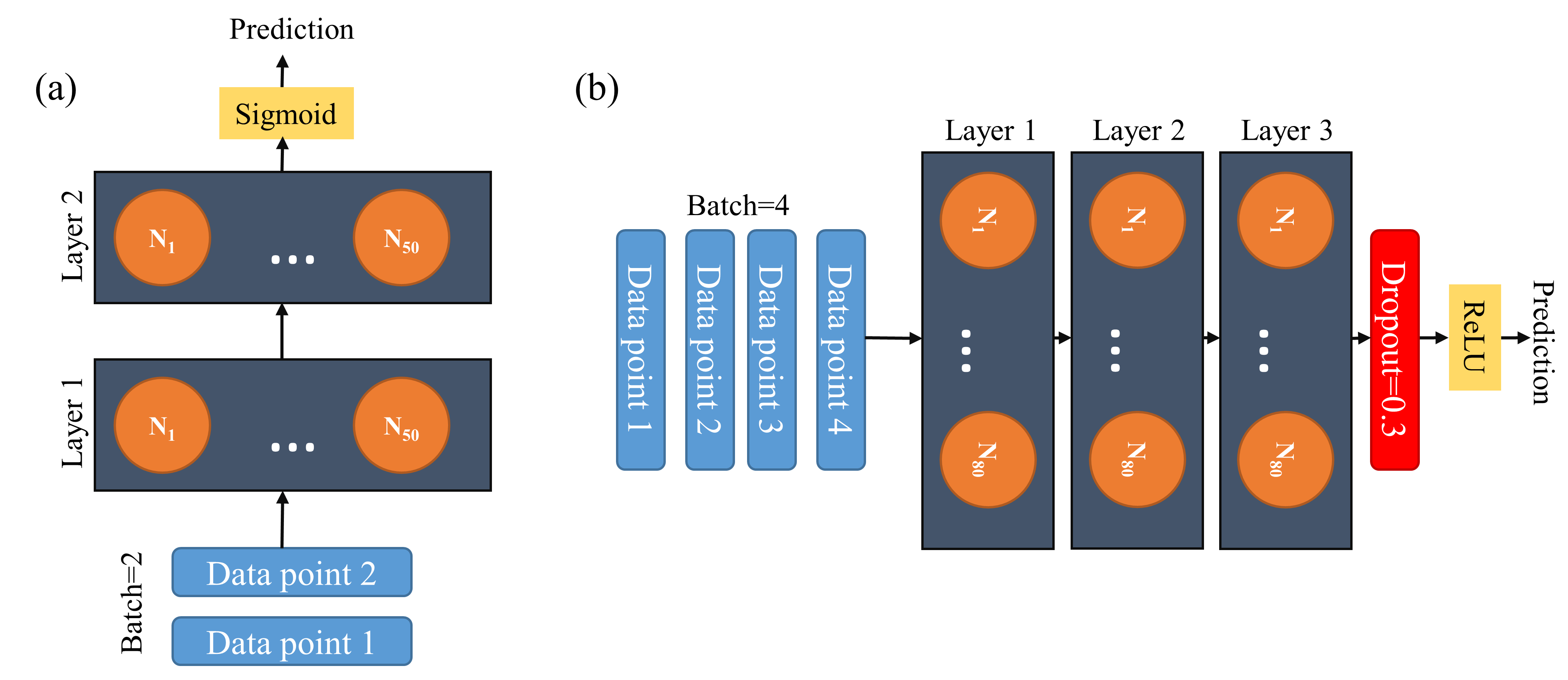}
    \caption{Examples of generated LSTMs. (\textbf{a}) LSTM with two hidden layers, 50 hidden units in each layer, batch size 2, and sigmoid activation function. (\textbf{b}) LSTM with three hidden layers, 80 hidden units in each layer, batch size 4, dropout 0.3, and ReLU activation function.}
  \label{fig:lstm2}
\end{figure}

{As the proposed solution generated different models for different data (countries), they had different numbers of layers, hidden units, activation functions, batch sizes, and learning rates, individually optimized for each one of them. Figure \ref{fig:lstm2} shows two examples of such LSTMs.}

\begin{figure}[H]
    \centering
    \includegraphics[width=\linewidth]{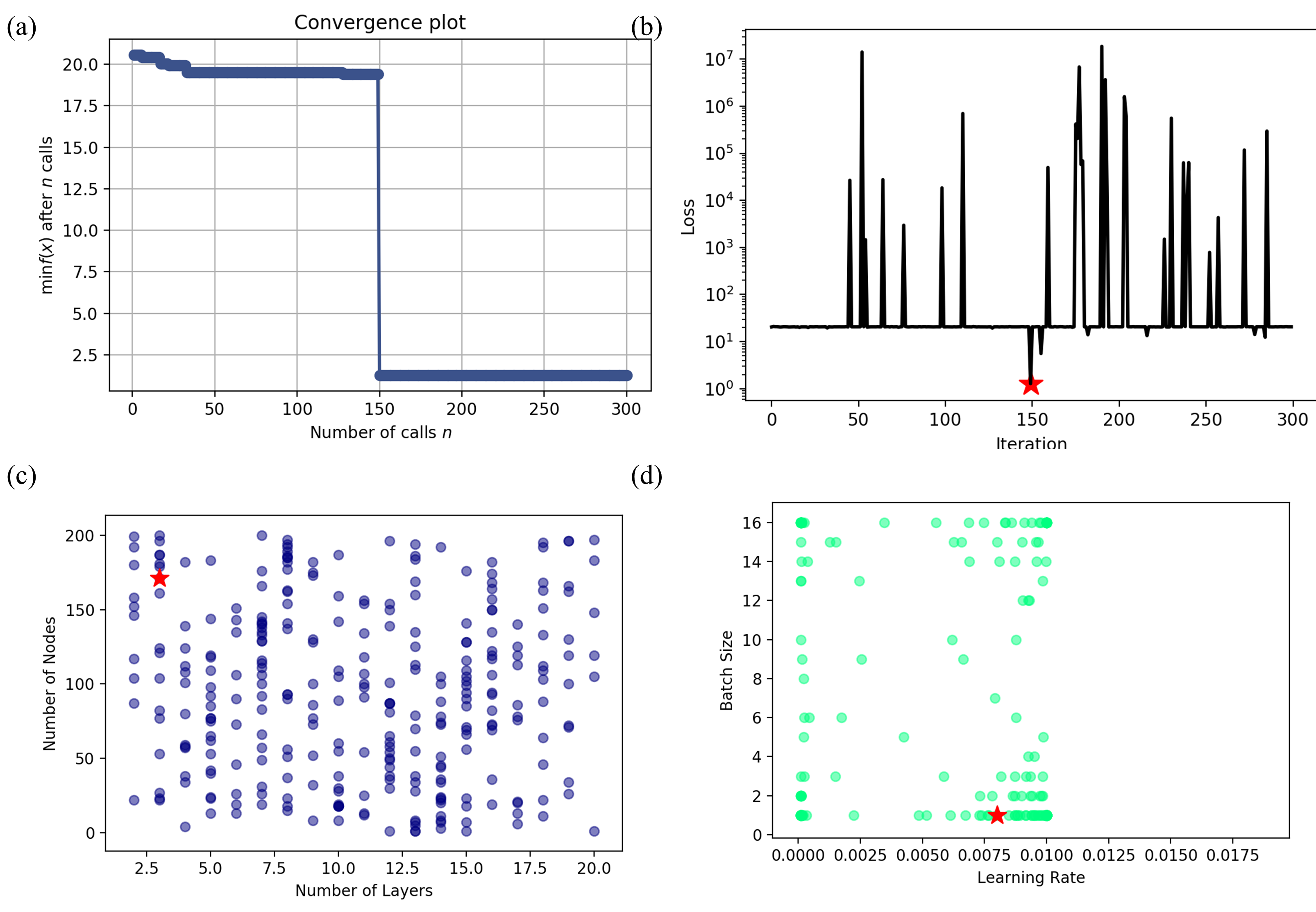}

    \caption{Results for the USA dataset. (\textbf{a}) minimum validation accuracy over iteration, (\textbf{b}) loss over iteration, distribution of (\textbf{c}) number of layers and hidden nodes, and (\textbf{d}) learning rates and batch size. (The red stars represent optimum value)}
    \label{fig:china1} \vspace{3mm} %MDPI: please confirm: whether these five-pointed star need explanations in the figures (b)-(d). Ans: Explanation added
\end{figure}

Case study (USA): Here we discuss the optimization output of the network trained for the USA dataset. The optimization ended with a network containing three hidden layers with 171 hidden nodes in each layer. The network's hyperparameters derived after optimization were as follows: learning rate 0.0002, zero dropout, batch size 6 and ReLU activation method. Figure~\ref{fig:china1}a showed minimum RMSE over iterations, (b) shows different RMSE over iterations. In (c), (d) the distribution of the number of layers and hidden units and the distribution of learning rate and batch size are shown, respectively.

\subsection{Training} 
\begin{figure}[H]
    \centering
    \includegraphics[width=\linewidth]{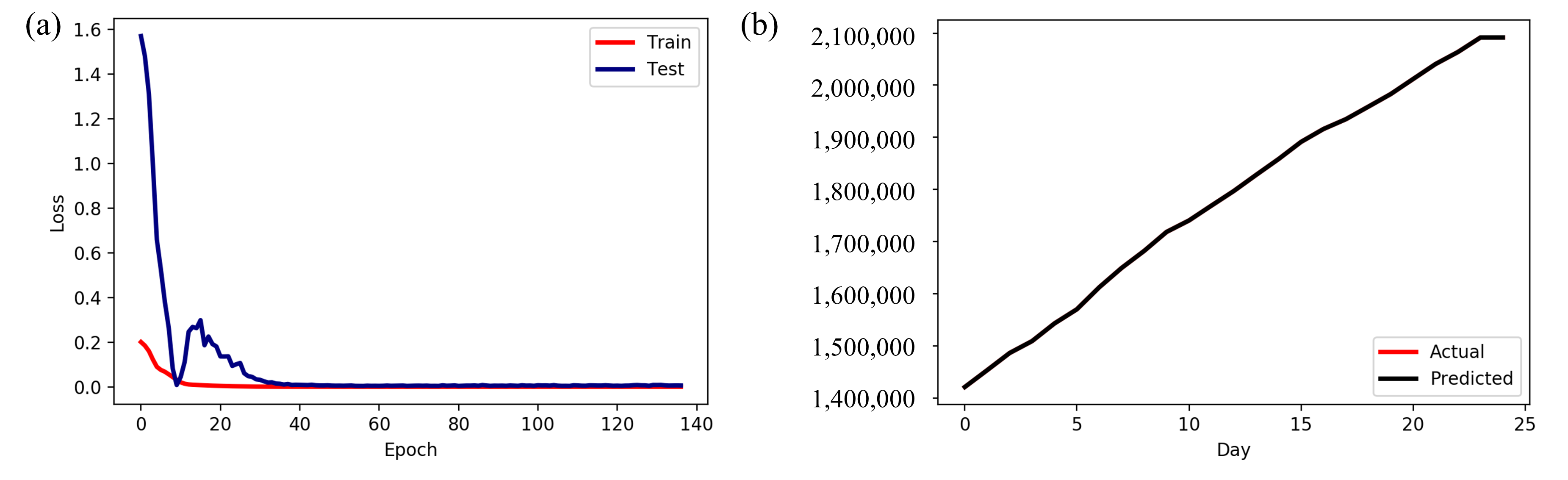}
    \caption{Case study of the network training for USA. (\textbf{a}) training loss during training, and (\textbf{b}) active case prediction on validation data.}
    \label{fig:china2} %MDPI: Please add the comma for the number which more than five digits in left subfigure, e.g., 16,000, 12,000, etc. or use scientific notation, the same to figure 9. Ans: Modified as suggested
\end{figure}

Each country-specific network was trained using its own case data. Although during optimization, the network was validated by predicting active cases, the same network was used to predict death, \mbox{recovery, and the} current number of cases. The networks were trained using a maximum of 5000 epochs combined with the early stopping mechanism used during optimization. The data of the last 25 days were used for test and the rest were used for training. Figure~\ref{fig:china2} shows the training loss over epochs and active case prediction using the network optimized for USA. It is noted that the loss lowered during training over epochs as it converged to a small value.

\subsection{Prediction Accuracy}
\begin{figure}[H]
    \centering
    \includegraphics[width=0.95\linewidth]{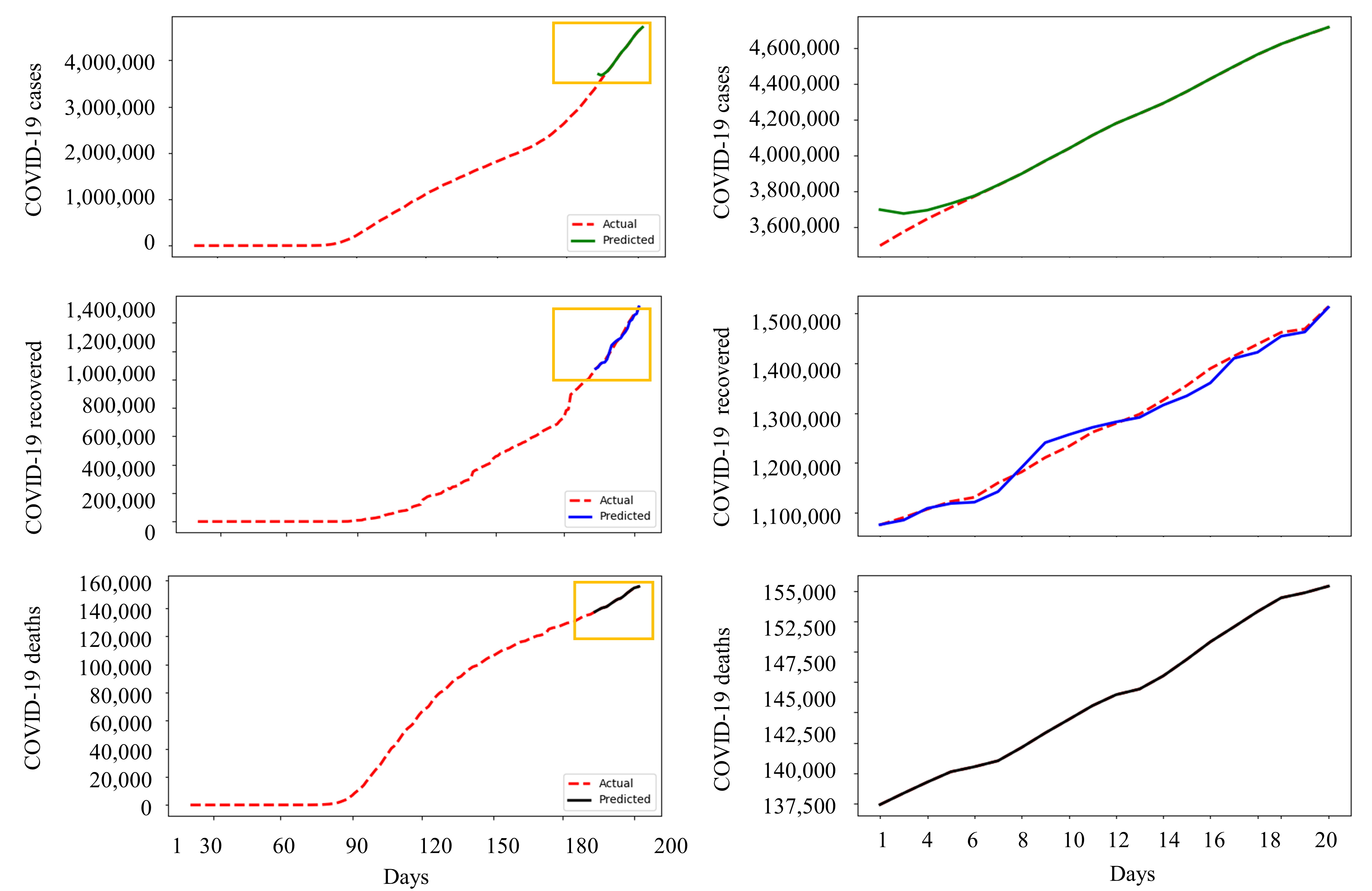}
    \caption{The 25 days ahead trend prediction in USA, COVID-19 cases (row 1), recovered (row 2), and deaths (row 3).}
    \label{fig:rmse} %MDPI: Please add the comma for the number which more than five digits in left subfigure, e.g., 16,000, 12,000, etc. or use scientific notation. Ans: Modified as suggested
\end{figure}

Here we discuss the prediction accuracy of the proposed method. The final fuzzy-rule based classification depended on death rate, {case rate}, and {recovery rate}. The suitable model chosen for each country was trained to predict these three values. We calculated root-mean-square error (RMSE) on the validation data to evaluate the methods. We compared using baseline algorithms such as linear regression, lasso linear regression, ridge regression. A single model was used to predict the values of all the countries. It is observed that such methods performed very poorly due to the small dataset. We also compared the method with some advanced neural networks such as a variation of LSTM combined with a fully convolutional network~\cite{karim2019multivariate}, a variation of residual RNN~\cite{goel2017r2n2}, and GRU~\cite{althelaya2018stock}. It is also noted that very deep networks also failed to predict accurately using such a small dataset. Bayesian optimized shallow GRU performs closer to our method. The results are summarized in Table~\ref{tab:res}. Example results on USA trend prediction are shown in Figure ~\ref{fig:rmse} for numbers of active cases, recovered cases and deaths. %is italic neccessary? the same to following format like this. Ans: removed

\begin{table}[H]
\caption{Average root-mean-square error (RMSE) of the last 25 days' prediction of numbers of active, recovered and death cases.}
\centering\scalebox{0.95}[0.95]{
\begin{tabular}{cccc}
\toprule
                                  & \multicolumn{3}{c}{\textbf{RMSE}}                                                                                                   \\ \cmidrule{2-4}
\multirow{-2}{*}{\textbf{Method}} & \textbf{COVID-19 Cases} & \textbf{Recovered} & \textbf{Death} \\ \noalign{\hrule height 0.5pt}
Liner Regression  & 2705.6 & 856.6 & 427.2 \\\midrule
Lasso Linear Regression  & 1905.9 & 333.4 & 175.3 \\\midrule
Ridge Regression & 2307.5 & 614.2 & 213.3 \\\midrule
Elastic Net~\cite{hans2011elastic} & 2307.3 & 2105.9 & 300.2 \\\midrule
LSTM-FCNS~\cite{karim2019multivariate} & 2605.4 & 1305.3 & 269.5 \\\midrule
Recidual RNN~\cite{goel2017r2n2} & 2905.4 & 1109.5 & 242.3 \\\midrule
GRU~\cite{althelaya2018stock} & 2605.3 & 923.0 & 163.5 \\\midrule
GRU+Baysian & 1275.0 & 422.9 & 100.2 \\\midrule
Proposed & 1103.5 & 329.0 & 101.9 \\\bottomrule
\end{tabular}}
\label{tab:res}
\end{table}

\subsection{Risk Classification Accuracy} 
\begin{figure}[H]
    \centering
    \includegraphics[width=0.55\linewidth]{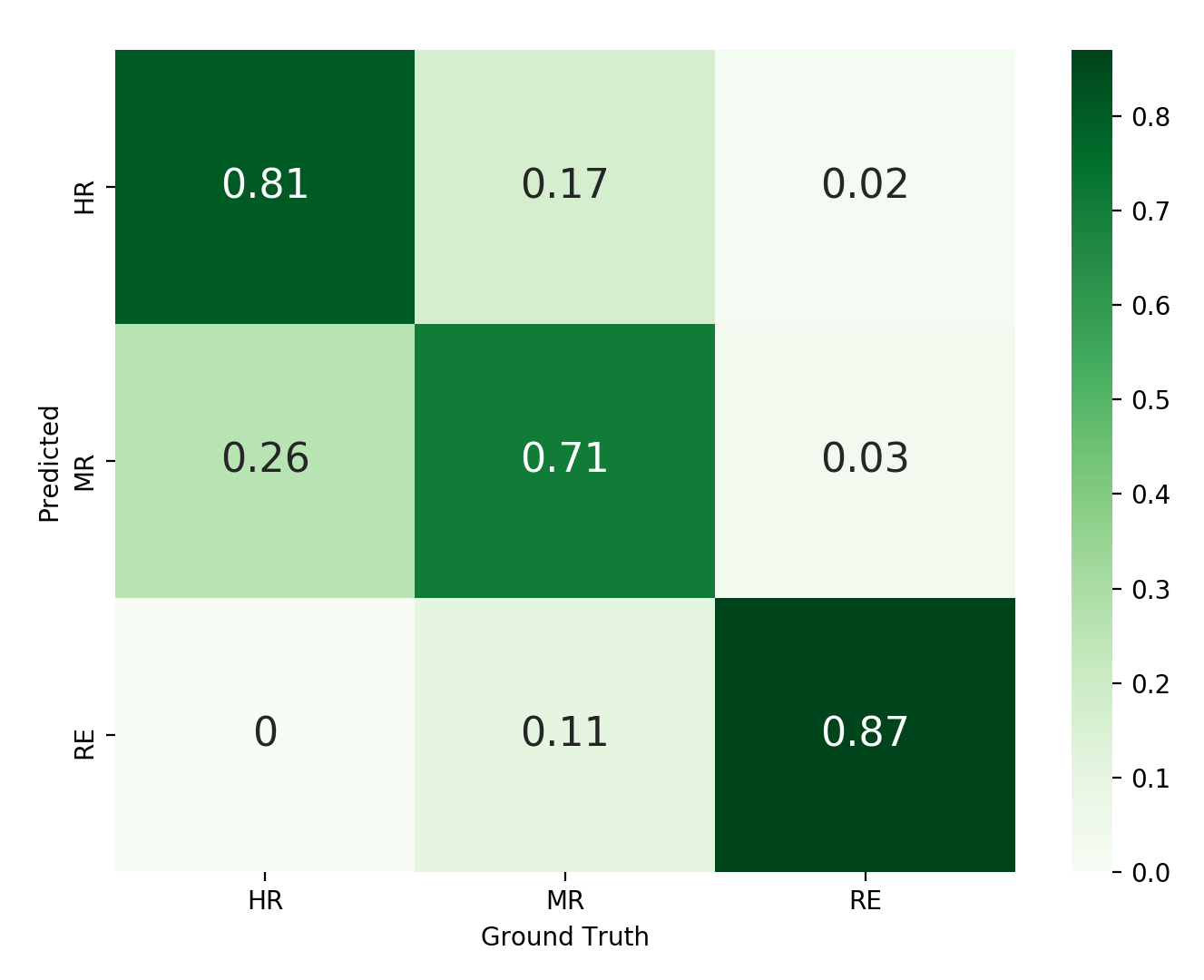}
    \caption{Confusion matrix of the four classes for 25 days ahead risk prediction for 180 countries.} %MDPI: Is color neccessary? if it is, please provide explanations. Ans: Color bar is give in the right side
   \label{fig:risk}
\end{figure}

A fuzzy rule-based method was used to classify the risk of each country into three classes as discussed earlier (HR, MR, and RE). We predicted the risk classes for 25 days ahead. The accuracy was calculated in a state-of-the-art manner using a manual ground truth extracted from the trend data. Figure~\ref{fig:risk} shows the confusion matrix of the classification accuracy over 180 countries. It is observed that the method produced relatively lower accuracy of predicting MR class due to the incorrect trend rate prediction. We achieved 77.6\% average accuracy over all the country-specific datasets. Table~\ref{tab:risk}~lists a few countries that are classified as high risk, low, and in recovering stage based on the prediction of August 2020.

\begin{table}[H]
\centering
\caption{Examples of some high risk, low risk, and recovering countries based on upcoming 25 days prediction (August'2020).}
\begin{tabular}{ccc}
\noalign{\hrule height 1.0pt}
\cellcolor[HTML]{FE0000}\textbf{\textcolor{white}{High Risk (HR)}} & \cellcolor[HTML]{3531FF}\textbf{\textcolor{white}{Low Risk (LR)}} & \cellcolor[HTML]{32CB00}\textbf{\textcolor{white}{Recovering (RE)}} \\\noalign{\hrule height 0.5pt}
                               India & Slovakia   & Greenland \\ \midrule
                               USA & Malta   & Uruguay \\ \midrule
                               Russia & Denmark   & Zimbabwe \\ \midrule
                               Brazil & Switzerland   & Japan \\ \midrule
                               Mexico & Germany   & Norway \\ \midrule

\end{tabular}
\label{tab:risk}
\end{table}

\subsection{{Implementation and Computational Cost}}
{The method was implemented using Python 3.6 combined in Anaconda environment. The Baysian optimization was implemented using open-source scikit-optimize library. The fitness function took the set of hyperparameters as the input and the validation loss was considered as faintness value. Minimum~loss was considered as the most fitted model.
Finally, the optimized properties of the network was used to design the LSTM using Tensorflow framework. The fuzzy rules were implemented using simple if-then rules.}
All the experiments were carried out in Intel(R) Xeon(R) Gold
6154 CPU with 128 GB of RAM and NVIDIA Quadro RTX
6000 GPU of capacity 24 GB. The method utilizes $\sim$72 computational hours for feature selection, network optimization, training, and evaluating the method.

\subsection{The Effect of Weather}
\begin{figure}[H]
    \centering
    \includegraphics[width=\linewidth]{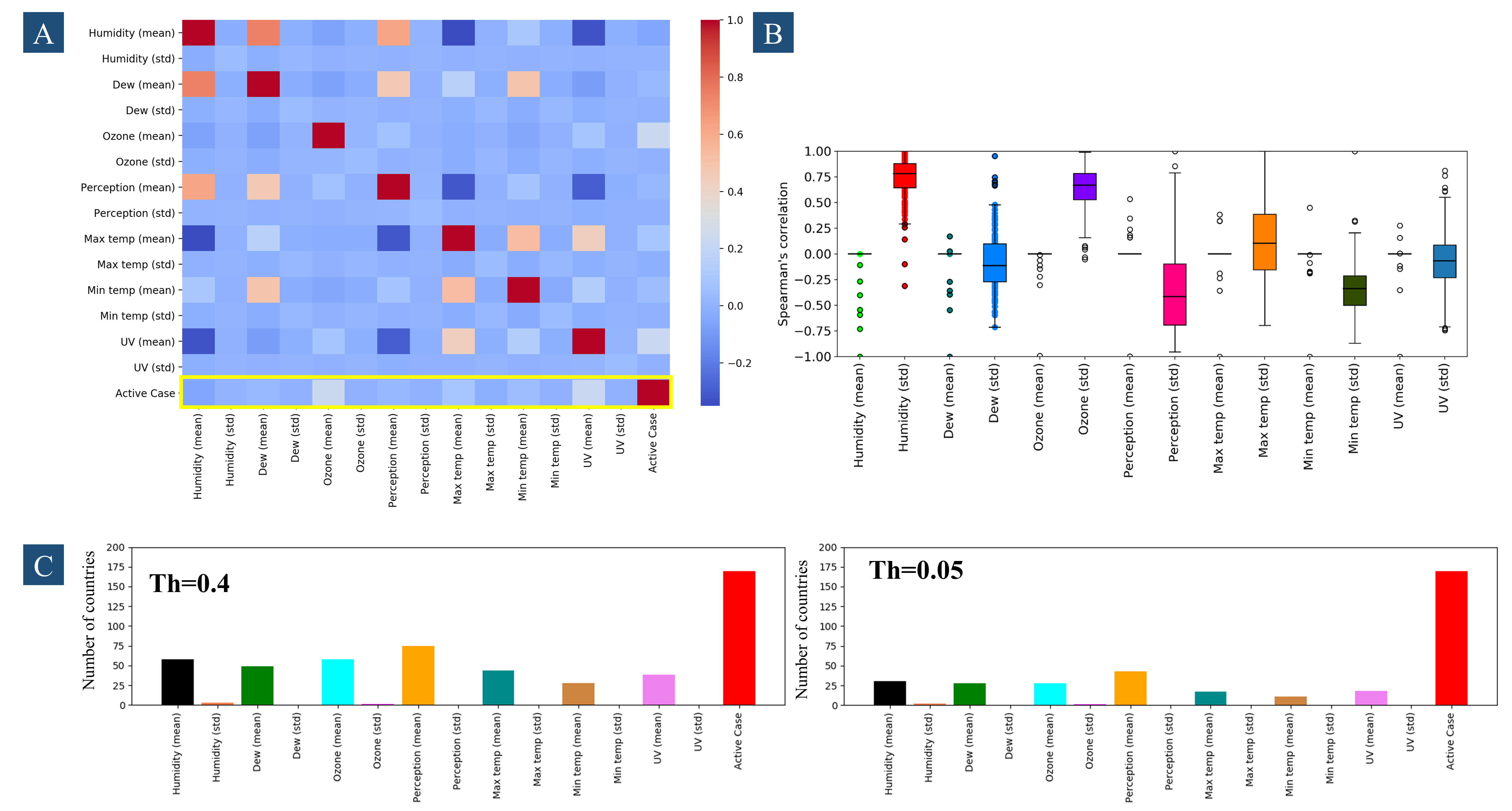}
    \caption{(\textbf{A}) Pearson correlation among features, the last row (yellow border) shows correlation with active case, (\textbf{B}) Distribution of the correlation considering all the countries, 
     (\textbf{C} Selected features by countries based on different thresholds.} %MDPI: please confirm: there is no explanations of subfigure C. Ans: Added
    \label{fig:cor}
\end{figure}

\begin{figure}[H]
    \centering
    \includegraphics[width=0.65\linewidth]{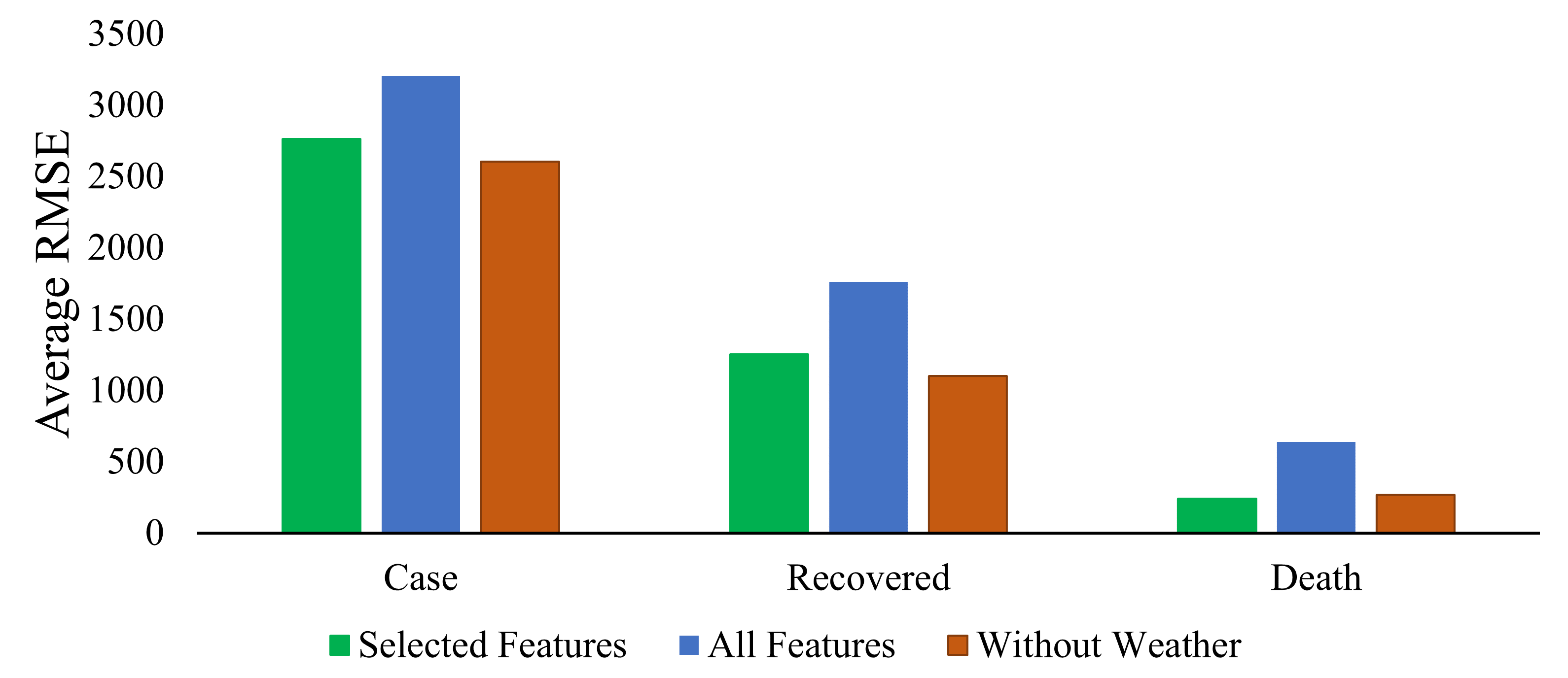}    
    \caption{Average RMSE of total cases, recovered, and death using selected features, combined~weather data, and without weather data on last 25 days prediction.}
    \label{fig:com}
\end{figure}

To understand the effect of weather, we experimented with combining weather data. The~weather data consisted of UV, minimum temperature, maximum temperature, perception, ozone, dew, and humidity. We took the mean and standard deviation for each feature. First, we combined the trend data and weather data and apply the network optimization method. Next, the optimized network was used to train and validate the results. We have verified the effect by (A) extracting the Pearson correlation coefficient among variables, (B) feature selection using ordinary least squares (OLS) regression from the combined data and finding optimized neural network, and (C) using all the features in combined data and applying the proposed method. Figure~\ref{fig:cor}A shows the average Pearson correlation of all the countries among different weather data and trend. It is observed that the trend data (active case) were not significantly correlated with the different weather parameters (see the last row, i.e., yellow border). Figure~\ref{fig:cor}B shows the distribution of Pearson correlation over different countries. It is noted that the weather parameters were not correlated or varied significantly.  It may be possible that all the features were not linked with the prediction variable. The data contained three main concerns for the risk categorization of a country. Number of cases ($\kappa$), number of deaths ($\delta$), and number of recoveries ($\rho$). The number of active cases ($\alpha$) was calculated by $\kappa-(\rho + \delta)$. Features were selected by backward elimination method. We calculated the p-value of all features with $\alpha$ using ordinary least squares (OLS) regression. We employ a threshold (0.4) for choosing features. Algorithm~\ref{alg:fs} demonstrates the method. Figure~\ref{fig:cor}C shows the number of times the features are selected by the countries. It is observed that choosing a lower threshold (bottom row) almost discarded all the weather data by most of the countries. Figure~\ref{fig:com} shows the average accuracy using the proposed OLS-based feature selection, combined~weather features, and without weather data. It is noted that the prediction accuracy was almost similar with and without weather data.  %MMDPI: is the bold necessary? if not, please remove it.
\vspace{6pt}

\begin{algorithm}[H]
\SetAlgoLined
\SetKwInOut{Input}{Input}
    \SetKwInOut{Output}{Output}
    \SetKwProg{myalg}{Procedure}{}{}
    \Input{A set of feature ($f$)}
    \Output{Selected set of feature ($\hat{f}$)}
\myalg{Feature Selection (f=set of features)}{
  Set $\hat{f}=f;$\\
  Set $p_{threshold}=0.4$\;
 \Do{$p_{max} >  p_{threshold}$}{
      Train OLS with $\hat{f}$\;
      $p_{max}= \forall \hat{f}$, MAX(p-value)\;
      \uIf{$p_{max} >  p_{threshold}$}{
            Remove $\hat{f}(p_{max})$ \;
        }

    }
    \Return $\hat{f}$
    }
 \caption{Feature selection algorithms}
 \label{alg:fs}
\end{algorithm}

\section{Conclusions}
In this paper, we have proposed a Bayesian optimization guided shallow LSTM for predicting the country-specific risk of the novel coronavirus (COVID-19). We have used the trend data to predict different parameters for the risk classification task. We also propose to use country-specific optimized network for accurate prediction and noted that this is suitable when we have a small and uncertain dataset. Combining the overall optimized LSTMs, we also note that a shallow network performs better compared to a deep neural networks. The method can be useful to predict the long-duration risk of an epidemic like COVID-19. We have also analyzed the prediction performance combining the weather data. We observed that the prediction model performs similarly without the weather data. We have shared the dataset used in this work at the project page (https://covid19prediction.github.io/)

In the future, we plan to explore a combination of different modalities of data such as flight, travelers, business, tourists, etc. The method can also be used to predict the economical effects of such~epidemics.

%%%%%%%%%%%%%%%%%%%%%%%%%%%%%%%%%%%%%%%%%%
\vspace{6pt}

%%%%%%%%%%%%%%%%%%%%%%%%%%%%%%%%%%%%%%%%%%
%% optional
%\supplementary{The following are available online at \linksupplementary{s1}, Figure S1: title, Table S1: title, Video S1: title.}

% Only for the journal Methods and Protocols:
% If you wish to submit a video article, please do so with any other supplementary material.
% \supplementary{The following are available at \linksupplementary{s1}, Figure S1: title, Table S1: title, Video S1: title. A supporting video article is available at doi: link.}

%%%%%%%%%%%%%%%%%%%%%%%%%%%%%%%%%%%%%%%%%%
\authorcontributions{conceptualization, R.P. and A.A.S.; methodology, R.P. and D.K.P.; software, R.P.; validation, S.K., D.K.P. and A.A.S.; formal analysis, investigation, resources, and data curation, R.P.; writing---original draft preparation, R.P.; writing---review and editing, A.A.S., D.K.P., and S.K.; visualization, A.A.S.; supervision, A.A.S. All authors have read and agreed to the published version of the manuscript.
}

%%%%%%%%%%%%%%%%%%%%%%%%%%%%%%%%%%%%%%%%%%
\funding{There is no funding reported for the research. The resources and publication supports are provided by UiT The Arctic University of Norway.}

%%%%%%%%%%%%%%%%%%%%%%%%%%%%%%%%%%%%%%%%%%
%\acknowledgments{\hl{xxxxx}}
%\acknowledgments{In this section you can acknowledge any support given which is not covered by the author contribution or funding sections. This may include administrative and technical support, or donations in kind (e.g., materials used for experiments).}

%%%%%%%%%%%%%%%%%%%%%%%%%%%%%%%%%%%%%%%%%%
\conflictsofinterest{The authors declare no conflict of interest.}

%%%%%%%%%%%%%%%%%%%%%%%%%%%%%%%%%%%%%%%%%%
%% optional

%%%%%%%%%%%%%%%%%%%%%%%%%%%%%%%%%%%%%%%%%%
% Citations and References in Supplementary files are permitted provided that they also appear in the reference list here.

%=====================================
% References, variant A: internal bibliography
%=====================================
\reftitle{References}
%\begin{thebibliography}{999}

%\end{thebibliography}

% The following MDPI journals use author-date citation: Arts, Econometrics, Economies, Genealogy, Humanities, IJFS, JRFM, Laws, Religions, Risks, Social Sciences. For those journals, please follow the formatting guidelines on http://www.mdpi.com/authors/references
% To cite two works by the same author: \citeauthor{ref-journal-1a} (\citeyear{ref-journal-1a}, \citeyear{ref-journal-1b}). This produces: Whittaker (1967, 1975)
% To cite two works by the same author with specific pages: \citeauthor{ref-journal-3a} (\citeyear{ref-journal-3a}, p. 328; \citeyear{ref-journal-3b}, p.475). This produces: Wong (1999, p. 328; 2000, p. 475)

%=====================================
% References, variant B: external bibliography
%=====================================

%%%%%%%%%%%%%%%%%%%%%%%%%%%%%%%%%%%%%%%%%%
%% optional
%\sampleavailability{Samples of the compounds ...... are available from the authors.}

%% for journal Sci
%\reviewreports{\\
%Reviewer 1 comments and authors’ response\\
%Reviewer 2 comments and authors’ response\\
%Reviewer 3 comments and authors’ response
%}

%%%%%%%%%%%%%%%%%%%%%%%%%%%%%%%%%%%%%%%%%%

\begin{thebibliography}{-------}
\providecommand{\natexlab}[1]{#1}

\end{thebibliography}


\begin{thebibliography}{999}
\providecommand{\natexlab}[1]{#1}

\bibitem[Wu \em{et~al.}(2020)Wu, Zhao, Yu, Chen, Wang, Song, Hu, Tao, Tian,
  Pei, et~al.]{wu2020new}
Wu, F.; Zhao, S.; Yu, B.; Chen, Y.M.; Wang, W.; Song, Z.G.; Hu, Y.; Tao, Z.W.;
  Tian, J.H.; Pei, Y.Y.; et al.
\newblock A new coronavirus associated with human respiratory disease in China.
\newblock {\em Nature} {\bf 2020}, {\em 579},~265--269.

\bibitem[Cheong and Jones(2020)]{cheongintroducing}
Cheong, K.H.; Jones, M.C.
\newblock Introducing the 21st Century's New Four Horsemen of the
  Coronapocalypse.
\newblock {\em BioEssays} {\bf 2020},  2000063, doi:10.1002/bies.202000063.

\bibitem[Akita \em{et~al.}(2016)Akita, Yoshihara, Matsubara, and
  Uehara]{akita2016deep}
Akita, R.; Yoshihara, A.; Matsubara, T.; Uehara, K.
\newblock Deep learning for stock prediction using numerical and textual
  information.
\newblock In Proceedings of the  2016 IEEE/ACIS 15th International Conference on Computer and
  Information Science (ICIS), Okayama, Japan, 26--29 June  2016; pp. 1--6.

\bibitem[Ali and Lee(2018)]{ali2018crm}
Ali, M.; Lee, Y.
\newblock CRM Sales Prediction Using Continuous Time-Evolving Classification.
\newblock In Proceedings of the  Thirty-Second AAAI Conference on Artificial Intelligence, New Orleans, LA, USA, 2--7 February 2018.

\bibitem[Xiao and Ai(2018)]{xiao2018data}
Xiao, F.; Ai, Q.
\newblock Data-driven multi-hidden markov model-based power quality disturbance
  prediction that incorporates weather conditions.
\newblock {\em IEEE Trans. Power Syst.} {\bf 2018}, {\em
  34},~402--412.

\bibitem[Lu \em{et~al.}(2019)Lu, Wang, Wang, Zhou, Zhang, Zhou, Niu, Chen, and
  Chou]{lu2019epidemic}
Lu, Y.; Wang, S.; Wang, J.; Zhou, G.; Zhang, Q.; Zhou, X.; Niu, B.; Chen, Q.;
  Chou, K.C.
\newblock An epidemic avian influenza prediction model based on google trends.
\newblock {\em Lett. Org. Chem.} {\bf 2019}, {\em 16},~303--310.

\bibitem[Lin \em{et~al.}(2020)Lin, Ho, Cheong, Li, Cai, Chee, Ng, Xiao, and
  Ong]{lin2020leveraging}
Lin, A.X.; Ho, A.F.W.; Cheong, K.H.; Li, Z.; Cai, W.; Chee, M.L.; Ng, Y.Y.;
  Xiao, X.; Ong, M.E.H.
\newblock Leveraging~Machine Learning Techniques and Engineering of
  Multi-Nature Features for National Daily Regional Ambulance Demand
  Prediction.
\newblock {\em Int. J. Environ. Res. Public Health} {\bf 2020}, {\em 17},~4179.

\bibitem[Tosepu \em{et~al.}(2020)Tosepu, Gunawan, Effendy, Lestari, Bahar,
  Asfian, et~al.]{tosepu2020correlation}
Tosepu, R.; Gunawan, J.; Effendy, D.S.; Lestari, H.; Bahar, H.; Asfian, P.
\newblock Correlation between weather and Covid-19 pandemic in Jakarta,
  Indonesia.
\newblock {\em Sci. Total. Environ.} {\bf 2020}, \emph{725}, 138436.

\bibitem[Gupta \em{et~al.}(2020)Gupta, Raghuwanshi, and
  Chanda]{gupta2020effect}
Gupta, S.; Raghuwanshi, G.S.; Chanda, A.
\newblock Effect of weather on COVID-19 spread in the US: A prediction model
  for India in 2020.
\newblock {\em Sci. Total. Environ.} {\bf 2020}, \emph{728}, 138860.

\bibitem[{\c{S}}ahin(2020)]{csahin2020impact}
{\c{S}}ahin, M.
\newblock Impact of weather on COVID-19 pandemic in Turkey.
\newblock {\em Sci. Total. Environ.} {\bf 2020}, \emph{728}, 138810.

\bibitem[Jia \em{et~al.}(2019)Jia, Li, Tan, and Xie]{jia2019predicting}
Jia, W.; Li, X.; Tan, K.; Xie, G.
\newblock Predicting the outbreak of the hand-foot-mouth diseases in China
  using recurrent neural network.
\newblock In Proceedings of the  2019 IEEE International Conference on Healthcare Informatics (ICHI), Xi'an, China, 10--13 June  2019; pp. 1--4.

\bibitem[Hamer \em{et~al.}(2020)Hamer, Birr, Verreet, Duttmann, and
  Klink]{hamer2020spatio}
Hamer, W.B.; Birr, T.; Verreet, J.A.; Duttmann, R.; Klink, H.
\newblock Spatio-Temporal Prediction of the Epidemic Spread of Dangerous
  Pathogens Using Machine Learning Methods.
\newblock {\em ISPRS Int. J. Geo-Inf.} {\bf 2020}, {\em
  9},~44.

\bibitem[Mezzatesta \em{et~al.}(2019)Mezzatesta, Torino, De~Meo, Fiumara, and
  Vilasi]{mezzatesta2019machine}
Mezzatesta, S.; Torino, C.; De~Meo, P.; Fiumara, G.; Vilasi, A.
\newblock A machine learning-based approach for predicting the outbreak of
  cardiovascular diseases in patients on dialysis.
\newblock {\em Comput. Methods Programs Biomed.} {\bf 2019}, {\em
  177},~9--15.

\bibitem[Jhuo \em{et~al.}(2019)Jhuo, Hsieh, Weng, Chen, Yang, and
  Yeh]{jhuo2019trend}
Jhuo, S.L.; Hsieh, M.T.; Weng, T.C.; Chen, M.J.; Yang, C.M.; Yeh, C.H.
\newblock Trend Prediction of Influenza and the Associated Pneumonia in Taiwan
  Using Machine Learning.
\newblock In Proceedings of the  2019 International Symposium on Intelligent Signal Processing and
  Communication Systems (ISPACS), Taipei, Taiwan, 3--6~December  2019;  pp. 1--2.

\bibitem[Kumar \em{et~al.}(2020)Kumar, Suresh, Reddy, and
  Reddy]{kumar2020outbreak}
Kumar, S.V.; Suresh, V.; Reddy, B.D.K.; Reddy, Y.J.
\newblock Outbreak Predictions in Healthcare Domain using Machine learning \&
  Artificial Intelligence.
\newblock {\em TEST Eng. Manag.} {\bf 2020}, {\em
  82},~11395--11400.

\bibitem[Machado \em{et~al.}(2019)Machado, Vilalta, Recamonde-Mendoza, Corzo,
  Torremorell, Perez, and VanderWaal]{machado2019identifying}
Machado, G.; Vilalta, C.; Recamonde-Mendoza, M.; Corzo, C.; Torremorell, M.;
  Perez, A.; VanderWaal,~K.
\newblock Identifying outbreaks of Porcine Epidemic Diarrhea virus through
  animal movements and spatial neighborhoods.
\newblock {\em Sci. Rep.} {\bf 2019}, {\em 9},~1--12.

\bibitem[Philemon \em{et~al.}(2019)Philemon, Ismail, and
  Dare]{philemon2019review}
Philemon, M.D.; Ismail, Z.; Dare, J.
\newblock A Review of Epidemic Forecasting Using Artificial Neural Networks.
\newblock {\em Int. J. Epidemiol. Res.} {\bf 2019},
  {\em 6},~132--143.

\bibitem[Abdulkareem \em{et~al.}(2020)Abdulkareem, Augustijn, Filatova, Musial,
  and Mustafa]{abdulkareem2020risk}
\textls[-15]{Abdulkareem, S.A.; Augustijn, E.W.; Filatova, T.; Musial, K.; Mustafa, Y.T.
\newblock Risk perception and behavioral change during epidemics: Comparing
  models of individual and collective learning.
\newblock {\em PLoS ONE} {\bf 2020}, \emph{15},~e0226483}.

\bibitem[Forna \em{et~al.}(2019)Forna, Nouvellet, Dorigatti, and
  Donnelly]{forna2019case}
Forna, A.; Nouvellet, P.; Dorigatti, I.; Donnelly, C.
\newblock Case fatality ratio estimates for the 2013--2016 West African Ebola
  epidemic: Application of Boosted Regression Trees for imputation.
\newblock {\em Int. J. Infect. Dis.} {\bf 2019}, {\em
  79},~128.

\bibitem[Dallatomasina \em{et~al.}(2015)Dallatomasina, Crestani,
  Sylvester~Squire, Declerk, Caleo, Wolz, Stinson, Patten, Brechard, Gbabai,
  et~al.]{dallatomasina2015ebola}
Dallatomasina, S.; Crestani, R.; Sylvester~Squire, J.; Declerk, H.; Caleo,
  G.M.; Wolz, A.; Stinson, K.; Patten, G.; Brechard, R.; Gbabai, O.B.M.;
  et al.
\newblock Ebola outbreak in rural West Africa: Epidemiology, clinical features
  and outcomes.
\newblock {\em Trop. Med. Int. Health} {\bf 2015}, {\em
  20},~448--454.

\bibitem[Plowright \em{et~al.}(2019)Plowright, Becker, Crowley, Washburne,
  Huang, Nameer, Gurley, and Han]{plowright2019prioritizing}
Plowright, R.K.; Becker, D.J.; Crowley, D.E.; Washburne, A.D.; Huang, T.;
  Nameer, P.; Gurley, E.S.; Han, B.A.
\newblock Prioritizing surveillance of Nipah virus in India.
\newblock {\em PLoS Negl. Trop. Dis.} {\bf 2019}, {\em 13}, e0007393.

\bibitem[Seetah \em{et~al.}(2020)Seetah, LaBeaud, Kumm, Grossi-Soyster,
  Anangwe, and Barry]{seetah2020archaeology}
Seetah, K.; LaBeaud, D.; Kumm, J.; Grossi-Soyster, E.; Anangwe, A.; Barry, M.
\newblock Archaeology and contemporary emerging zoonosis: A framework for
  predicting future Rift Valley fever virus outbreaks.
\newblock {\em Int. J. Osteoarchaeol.} {\bf 2020}, doi:10.1002/oa.2862.

\bibitem[Rao and Vazquez(2020)]{rao2020identification}
Rao, A.S.S.; Vazquez, J.A.
\newblock Identification of COVID-19 Can be Quicker through Artificial
  Intelligence framework using a Mobile Phone-Based Survey in the Populations
  when Cities/Towns Are Under Quarantine.
\newblock {\em Infect. Control. Hosp. Epidemiol.} {\bf 2020}, \emph{41}, 826--830.

\bibitem[Yan \em{et~al.}(2020)Yan, Zhang, Xiao, Wang, Sun, Liang, Li, Zhang,
  Guo, Xiao, et~al.]{yan2020prediction}
Yan, L.; Zhang, H.T.; Xiao, Y.; Wang, M.; Sun, C.; Liang, J.; Li, S.; Zhang,
  M.; Guo, Y.; Xiao,~Y.;~et al.~Prediction~of criticality in patients with severe Covid-19 infection
  using three clinical features: A machine learning-based prognostic model with
  clinical data in Wuhan.~{\em medRxiv} {\bf 2020}, doi:10.1101/2020.02.27.20028027.

\bibitem[Peng \em{et~al.}(2020)Peng, Yang, Zhang, Zhuge, and
  Hong]{peng2020epidemic}
Peng, L.; Yang, W.; Zhang, D.; Zhuge, C.; Hong, L.
\newblock Epidemic analysis of COVID-19 in China by dynamical modeling.
\newblock {\em arXiv} {\bf 2020}, arXiv:2002.06563.

\bibitem[Zhao \em{et~al.}(2020)Zhao, Lin, Ran, Musa, Yang, Wang, Lou, Gao,
  Yang, He, et~al.]{zhao2020preliminary}
Zhao, S.; Lin, Q.; Ran, J.; Musa, S.S.; Yang, G.; Wang, W.; Lou, Y.; Gao, D.;
  Yang, L.; He, D.; et al.
\newblock Preliminary estimation of the basic reproduction number of novel
  coronavirus (2019-nCoV) in China, from 2019 to 2020: A data-driven analysis
  in the early phase of the outbreak.
\newblock {\em Int. J. Infect. Dis.} {\bf 2020}, {\em
  92},~214--217.

\bibitem[Chen \em{et~al.}(2020)Chen, Shi, Ni, Ruan, Jiang, Yao, Wang, Song,
  Zhou, and Ge]{chen2020data}
Chen, B.; Shi, M.; Ni, X.; Ruan, L.; Jiang, H.; Yao, H.; Wang, M.; Song, Z.;
  Zhou, Q.; Ge, T.
\newblock Data Visualization Analysis and Simulation Prediction for COVID-19.
\newblock {\em arXiv} {\bf 2020}, arXiv:2002.07096.

\bibitem[Li \em{et~al.}(2020)Li, Chen, and Deng]{li2020scaling}
Li, M.; Chen, J.; Deng, Y.
\newblock Scaling features in the spreading of COVID-19.
\newblock {\em arXiv} {\bf 2020}, arXiv:2002.09199.

\bibitem[Hilton and Keeling(2020)]{hilton2020estimation}
Hilton, J.; Keeling, M.J.
\newblock Estimation of country-level basic reproductive ratios for novel
  Coronavirus (COVID-19) using synthetic contact matrices.
\newblock {\em medRxiv} {\bf 2020}, doi:2020.02.26.20028167 .

\bibitem[Kastner \em{et~al.}(2020)Kastner, Wei, and Samet]{kastner2020viewing}
Kastner, J.; Wei, H.; Samet, H.
\newblock Viewing the Progression of the Novel Corona Virus (COVID-19) with
  NewsStand.
\newblock {\em arXiv} {\bf 2020}, arXiv:2003.00107.

\bibitem[Jia \em{et~al.}(2020)Jia, Li, Jiang, Guo, et~al.]{jia2020prediction}
Jia, L.; Li, K.; Jiang, Y.; Guo, X.~Prediction and analysis of Coronavirus Disease 2019.~{\em arXiv} {\bf 2020}, arXiv:2003.05447.

\bibitem[Zhao \em{et~al.}(2020)Zhao, Liu, and Li]{zhao2020tracking}
Zhao, X.; Liu, X.; Li, X.
\newblock Tracking the spread of novel coronavirus (2019-nCoV) based on big
  data.
\newblock {\em medRxiv} {\bf 2020}, doi:10.1101/2020.02.07.20021196 .

\bibitem[Zeng \em{et~al.}(2020)Zeng, Zhang, Li, Liu, and
  Qiu]{zeng2020predictions}
Zeng, T.; Zhang, Y.; Li, Z.; Liu, X.; Qiu, B.
\newblock Predictions of 2019-ncov transmission ending via comprehensive
  methods.
\newblock {\em arXiv} {\bf 2020}, arXiv:2002.04945.

\bibitem[Buizza(2020)]{buizza2020probabilistic}
Buizza, R.
\newblock Probabilistic prediction of COVID-19 infections for China and Italy,
  using an ensemble of stochastically-perturbed logistic curves.
\newblock {\em arXiv} {\bf 2020}, arXiv:2003.06418.

\bibitem[Fong \em{et~al.}(2020)Fong, Li, Dey, Crespo, and
  Herrera-Viedma]{fong2020finding}
Fong, S.J.; Li, G.; Dey, N.; Crespo, R.G.; Herrera-Viedma, E.
\newblock Finding an accurate early forecasting model from small dataset: A
  case of 2019-ncov novel coronavirus outbreak.
\newblock {\em Int. J. Interact. Multimed. Artif. Intell.} {\bf 2020}, {\em 6},~51--61.

\bibitem[Santosh(2020)]{santosh2020ai}
Santosh, K.
\newblock AI-Driven Tools for Coronavirus Outbreak: Need of Active Learning and
  Cross-Population Train/Test Models on Multitudinal/Multimodal Data.
\newblock {\em J. Med Syst.} {\bf 2020}, {\em 44},~1--5.

\bibitem[Al-qaness \em{et~al.}(2020)Al-qaness, Ewees, Fan, and Abd
  El~Aziz]{al2020optimization}
Al-qaness, M.A.; Ewees, A.A.; Fan, H.; Abd El~Aziz, M.
\newblock Optimization Method for Forecasting Confirmed Cases of COVID-19 in
  China.
\newblock {\em J. Clin. Med.} {\bf 2020}, {\em 9},~674.

\bibitem[Li \em{et~al.}(2020)Li, Liang, Yin, Liu, Hao, Hu, Wang, and
  Jin]{li2020covid}
Li, Y.; Liang, M.; Yin, X.; Liu, X.; Hao, M.; Hu, Z.; Wang, Y.; Jin, L.
\newblock COVID-19 Epidemic Outside China: 34~Founders and Exponential Growth.
\newblock {\em medRxiv} {\bf 2020}, doi:10.1101/2020.03.01.20029819 .

\bibitem[Araujo and Naimi(2020)]{araujo2020spread}
Araujo, M.B.; Naimi, B.
\newblock Spread of SARS-CoV-2 Coronavirus likely to be constrained by climate.
\newblock {\em medRxiv} {\bf 2020}, doi:10.1101/2020.03.12.20034728.

\bibitem[Dong and Sun(2019)]{dong2019partial}
Dong, M.; Sun, J.
\newblock Partial Discharge Detection on Aerial Covered Conductors Using
  Time-Series Decomposition and Long Short-term Memory Network.
\newblock {\em arXiv} {\bf 2020}, arXiv:1907.03378.

\bibitem[Moews \em{et~al.}(2019)Moews, Herrmann, and Ibikunle]{moews2019lagged}
Moews, B.; Herrmann, J.M.; Ibikunle, G.
\newblock Lagged correlation-based deep learning for directional trend change
  prediction in financial time series.
\newblock {\em Expert Syst. Appl.} {\bf 2019}, {\em
  120},~197--206.

\bibitem[Thomas(2019)]{thomas2019time}
Thomas, K.
\newblock Time Series Prediction for Stock Price and Opioid Incident Location.
\newblock Ph.D. Thesis, Arizona State University: Tempe, AZ, USA,   2019.

\bibitem[Lorenzo and Olivas(2019)]{lorenzo2019some}
Lorenzo, A.; Olivas, J.A.
\newblock Some Considerations on the Use of AI Techniques for Prediction and
  Forecasting in Political Elections and Stock Market.
\newblock In Proceedings of the International Conference on Artificial
  Intelligence (ICAI), Jinan, China, 30  July–2  August 2019; Budapest, Hungary,  2019; pp.~403--407. %MDPI: please confirm the publisher name Ans: deleted
 %Author 1, A.B.; Author 2, C.D.; Author 3, E.F. Title of presentation. In Title of Collected Work, Proceedings of the Name of the Conference, Location of Conference, Country, Date of Conference (Day Month Year); Editor 1, Editor 2, Eds. (if available); Publisher: City, Country, Year; Abstract Number (optional), Pagination (optional).

 % or delete the highlight part.
 % Author 1, F.M.; Author 2, F.M.; Author 3, F.M.; et al. Title of presentation. In Proceedings of the Name of the Conference, Location of Conference, Country, Date of Conference (Day Month Year); Abstract Number (optional), Pagination (optional).

\bibitem[Bandara \em{et~al.}(2019)Bandara, Shi, Bergmeir, Hewamalage, Tran, and
  Seaman]{bandara2019sales}
Bandara, K.; Shi, P.; Bergmeir, C.; Hewamalage, H.; Tran, Q.; Seaman, B.
\newblock Sales demand forecast in e-commerce using a long short-term memory
  neural network methodology.~In Proceedings of the  International Conference on Neural Information Processing, Sydney, NSW, Australia, 12--15 December 2019; pp. 462--474.

\bibitem[Cui \em{et~al.}(2019)Cui, Hou, Sheng, Jiang, He, Jiang, Chi, and
  Tong]{cui2019prediction}
Cui, Y.; Hou, H.; Sheng, G.; Jiang, X.; He, M.; Jiang, G.; Chi, J.; Tong, J.
\newblock A prediction method for power transformer state parameters based on
  feature attention mechanism.
\newblock In Proceedings of the  2019 IEEE PES Asia-Pacific Power and Energy Engineering Conference
  (APPEEC), Macao, China, 1--4  December 2019; pp.~1--5.

\bibitem[Oh \em{et~al.}(2018)Oh, Ng, San~Tan, and Acharya]{oh2018automated}
Oh, S.L.; Ng, E.Y.; San~Tan, R.; Acharya, U.R.
\newblock Automated diagnosis of arrhythmia using combination of CNN and LSTM
  techniques with variable length heart beats.
\newblock {\em Comput. Biol. Med.} {\bf 2018}, {\em
  102},~278--287.

\bibitem[Ho \em{et~al.}(2019)Ho, To, Koh, and Cheong]{ho2019forecasting}
Ho, A.F.W.; To, B.Z.Y.S.; Koh, J.M.; Cheong, K.H.
\newblock Forecasting Hospital Emergency Department Patient Volume Using
  Internet Search Data.
\newblock {\em IEEE Access} {\bf 2019}, {\em 7},~93387--93395.

\bibitem[Jim{\'e}nez \em{et~al.}(2020)Jim{\'e}nez, Palma, S{\'a}nchez,
  Mar{\'\i}n, Palacios, and L{\'o}pez]{jimenez2020feature}
Jim{\'e}nez, F.; Palma, J.; S{\'a}nchez, G.; Mar{\'\i}n, D.; Palacios, F.;
  L{\'o}pez, L.
\newblock Feature Selection based Multivariate Time Series Forecasting: An
  Application to Antibiotic Resistance Outbreaks Prediction.
\newblock {\em Artif. Intell. Med.} {\bf 2020}, \emph{104},  101818.

\bibitem[Tapak \em{et~al.}(2019)Tapak, Hamidi, Fathian, and
  Karami]{tapak2019comparative}
Tapak, L.; Hamidi, O.; Fathian, M.; Karami, M.
\newblock Comparative evaluation of time series models for predicting influenza
  outbreaks: Application of influenza-like illness data from sentinel sites of
  healthcare centers in Iran.
\newblock {\em BMC Res. Notes} {\bf 2019}, {\em 12},~353.

\bibitem[Su \em{et~al.}(2019)Su, Xu, Li, Ruan, Li, Deng, Li, Li, Chen, Xiong,
  et~al.]{su2019forecasting}
Su, K.; Xu, L.; Li, G.; Ruan, X.; Li, X.; Deng, P.; Li, X.; Li, Q.; Chen, X.;
  Xiong, Y.; et al. Forecasting~influenza activity using self-adaptive AI model and
  multi-source data in Chongqing, China.~{\em EBioMedicine} {\bf 2019},~{\em 47},~284--292.

\bibitem[Ochodek \em{et~al.}(2020)Ochodek, Kopczy{\'n}ska, and
  Staron]{ochodek2020deep}
Ochodek, M.; Kopczy{\'n}ska, S.; Staron, M.
\newblock Deep learning model for end-to-end approximation of COSMIC functional
  size based on use-case names.
\newblock {\em Inf. Softw. Technol.} {\bf 2020}, \emph{123}, 106310.

\bibitem[Hu \em{et~al.}(2020)Hu, Zhu, Liu, and Li]{hu2020efficient}
Hu, F.; Zhu, Y.; Liu, J.; Li, L.
\newblock An efficient Long Short-Term Memory model based on Laplacian Eigenmap
  in artificial neural networks.
\newblock {\em Appl. Soft Comput.} {\bf 2020}, \emph{91}, 106218.

\bibitem[Wen \em{et~al.}(2019)Wen, Wang, Tang, Xu, Li, and Zhao]{wen2019real}
Wen, S.; Wang, Y.; Tang, Y.; Xu, Y.; Li, P.; Zhao, T.
\newblock Real-Time Identification of Power Fluctuations Based on LSTM
  Recurrent Neural Network: A Case Study on Singapore Power System.
\newblock {\em IEEE Trans. Ind. Inform.} {\bf 2019}, {\em
  15},~5266--5275.

\bibitem[Yuan \em{et~al.}(2019)Yuan, Wang, Lin, Liu, and Yu]{yuan2019novel}
Yuan, J.; Wang, H.; Lin, C.; Liu, D.; Yu, D.
\newblock A novel GRU-RNN network model for dynamic path planning of mobile
  robot.
\newblock {\em IEEE Access} {\bf 2019}, {\em 7},~15140--15151.

\bibitem[Karim \em{et~al.}(2019)Karim, Majumdar, Darabi, and
  Harford]{karim2019multivariate}
Karim, F.; Majumdar, S.; Darabi, H.; Harford, S.
\newblock Multivariate lstm-fcns for time series classification.
\newblock {\em Neural~Netw.} {\bf 2019}, {\em 116},~237--245.

\bibitem[Goel \em{et~al.}(2017)Goel, Melnyk, and Banerjee]{goel2017r2n2}
Goel, H.; Melnyk, I.; Banerjee, A.
\newblock R2N2: Residual recurrent neural networks for multivariate time series
  forecasting.
\newblock {\em arXiv} {\bf 2017}, arXiv:1709.03159.

\bibitem[Althelaya \em{et~al.}(2018)Althelaya, El-Alfy, and
  Mohammed]{althelaya2018stock}
Althelaya, K.A.; El-Alfy, E.S.M.; Mohammed, S.
\newblock Stock market forecast using multivariate analysis with bidirectional
  and stacked (LSTM, GRU).
\newblock In Proceedings of the  2018 21st Saudi Computer Society National Computer Conference (NCC), Riyadh, Saudi Arabia, 25--26 April  2018; pp. 1--7.

\bibitem[Hans(2011)]{hans2011elastic}
Hans, C.~Elastic net regression modeling with the orthant normal prior.~{\em J. Am. Stat. Assoc.} {\bf 2011},~{\em 106}, 1383--1393.

\end{thebibliography}
\end{document}